\documentclass{mn2e}
\usepackage{amssymb}
\usepackage{graphicx}
\newcommand{\apjl}{ApJ}

\newcommand{\nustar}{NuSTAR J095551+6940.8}
\newcommand{\titolo}{NuSTAR J095551+6940.8}
\begin{document}

\title[Accreting NS as ULX]
{NuSTAR J095551+6940.8: a highly-magnetised neutron star with super-Eddington mass accretion}

\author[S.~Dall'Osso et al.]{Simone Dall'Osso$^1$, Rosalba Perna$^2$, Luigi Stella$^3$\\
$^1$Theoretical Astrophysics, University of T\"{u}bingen, Auf der Morgenstelle 10, 72076, Germany\\
$^2$ Department of Physics and Astronomy, Stony Brook University, Stony Brook, NY, USA\\
$^3$ INAF - Osservatorio Astronomico di Roma, via di Frascati 33, 00044, Monteporzio Catone, Roma, Italy}

\maketitle
\label{firstpage}


\begin{abstract}
The identification of the Ultraluminous X-ray source (ULX) X-2 in M82
as an accreting pulsar has shed new light on the nature of a
subset of ULXs, while rising new questions on the nature of the
super-Eddington accretion. Here, by numerically solving the torque
equation of the accreting pulsar within the framework of the
magnetically threaded-disk scenario, we show that three classes of
solutions, corresponding to different values of the magnetic field, 
are mathematically allowed.  We argue that the highest magnetic field one, corresponding to $B\sim
10^{13}$~G, is favoured based on physical considerations and the observed
properties of the source.  In particular, that is the only solution
which can account for the observed variations in $\dot{P}$ (over four
time intervals) without requiring major changes in $\dot{M}$, which would be
at odds with the approximately constant X-ray emission of the source
during the same time.  For this solution, we find that the source can only
accomodate a moderate amount of beaming, $0.5 \lesssim b < 1$.  Last,
we show that the upper limit on the luminosity, $L_X <  2.5\times
10^{38}$~erg~s$^{-1}$ from archival observations, is consistent with
a highly-magnetized neutron star being in the propeller phase at that
time.

\end{abstract}


\section{Introduction}

Bright X-ray sources, not associated with
galactic nuclei, were discovered during surveys in the late 1970s and
early 1980s with the {\em ROSAT} and {\em Einstein} telescopes.  These
surveys showed that the X-ray luminosity function of typical galaxies
extends up to luminosities on the order of $\sim 10^{42}$~erg~s$^{-1}$
(e.g., Fabbiano 1989 for a review).
The contribution to the bright end of the luminosity distribution 
was revealed to be made up of individual sources thanks to the
superior spatial resolution and sensitivity of the {\em Chandra}
telescope (e.g. the case of the Antennae; Zesas et al. 2002). 
In addition, observations with {\em Chandra} also revealed that
some early type galaxies, while having no sign of nuclear activity, 
still displayed the presence of individual sources with 
luminosities in the $\sim 10^{41}-10^{42}$~erg~s$^{-1}$ range
 (e.g. Hornschemeier et al. 2004; Ptak \& Colbert 2004).

 These point-like sources, emitting X-rays at values exceeding the
 Eddington limit for a solar mass compact object, have been
 collectively dubbed ultraluminous X-ray sources (ULXs). Their
 physical nature has long been debated in the literature, with a
 variety of models and ideas put forward to explain their
 properties. These include, among others, accreting intermediate-mass
 black holes with masses $\sim 10^2-10^4 M_\odot$(e.g. Colbert \& Mushotzky 1999;
Miller et al. 2003), beamed X-ray binaries (King et al. 2001), young supernova
 remnants in extremely dense environments (Marston et al. 1995),
 super-Eddington accretors (e.g. Begelman et al. 2006), young, rotation-powered
 pulsars (Perna \& Stella 2004). It is possible that ULXs 
 form a heterogeneous family consisting of different sub-classes
 (e.g. Gladstone et al. 2013).

The recent detection of 1.37 s pulsations in the X-ray
flux of an ULX in the M82 galaxy has allowed to identify this source
as an accreting magnetized NS (Bachetti et al. 2014). The average period
derivative during five of the NuStar observations indicated that the
source is spinning up, as typical of a NS accreting matter from a
companion star.  The super-Eddington luminosity of the source could then be
accounted for by a combination of geometrical beaming and reduced
opacities due to a strong magnetic field. However, Bachetti et al. (2014)
highlighted a possible inconsistency between the value of the magnetic
field inferred from measurements of the period derivative and that
inferred from the X-ray luminosity, if due to accretion.

In this paper we carry out a detailed modeling of the timing
properties of the source; we show that, by applying the Ghosh \& Lamb 
(1979; GL in the following) model for magnetically-threaded disks
in its general formulation, both the period derivative and the X-ray luminosity 
of the source can be self-consistently accounted for. Our work complements 
and/or extends recent results which have appeared while we were working 
on our paper (Eksi et al. 2014; Lyutikov 2014; Kluzniak \& Lasota 2014; Tong 2014).
\section{Magnetic threading of accretion disks}
\label{sec:coupling}
The pulsations seen from the source and the measured spin-up indicate
that an accretion disk exists around the NS, and that it should be
truncated at some distance from the NS
where matter is forced to flow 
towards a localised region on the stellar surface. The NS magnetic field is
responsible for truncating the disk at the magnetospheric radius,
$r_m$, and for channeling the accretion flow along the B-field lines.

We can write 
the X-ray luminosity, L$_X$, in terms of the accretion luminosity, L$_{\rm acc}$, as
\begin{equation}
\label{eq:define-eta}
{\rm L}_{\rm X} \; = \; \eta {\rm L}_{\rm acc} \;= \;\eta \frac{G M}{R_{\rm ns}} \dot{M}\,,
\end{equation}
where $\dot{M}$ is the mas accretion rate and $\eta \sim$ unity is the efficiency 
of conversion of released accretion energy to X-rays. 
The ``isotropic" equivalent X-ray luminosity can then be expressed as 
L$_{X,{\rm iso}}$ =  L$_{\rm X}/b$ = L$_{\rm acc} (\eta/b)$,
where $b <1$ is the beaming factor of the emission. 

Adopting the value L$_{X,{\rm iso}} = 10^{40}$ erg s$^{-1}$
(Bachetti et al. 2014) we obtain
\begin{equation}
\label{eq:mdot}
\dot{m} = \frac{\dot{M}}{\dot{M}_{\rm E}} \approx 60~\frac{b}{\eta}
\left(\frac{L_{X,{\rm iso}}}{10^{40}~{\rm erg~s}^{-1}}\right) \, ,
\end{equation}

where the Eddington luminosity is\footnote{This expression holds for hydrogen.
A factor $A/Z$ should be added for a generic element of atomic
  weight $A$ and number $Z$.}  L$_{\rm E} = 4 \pi c G M m_p / \sigma_T
\simeq 1.75 \times 10^{38} (M/1.4 M_{\odot})$ erg s$^{-1}$, and
$\dot{M}_E = {\rm L}_E R_{\rm ns}/(G M) \simeq 1.41 \times
10^{-8} (R_{\rm ns}/10~{\rm km})$ M$_{\odot}$ yr$^{-1}$.

Finally  the disk luminosity, $L_{\rm disk}=(GM\dot{M})/2r_m$ is
\begin{equation}
\label{eq:disc-luminosity}
\frac{{\rm L}_{\rm disc}}{{\rm L}_{\rm E}} \simeq 0.3 
\left(\frac{\dot{m}}{60}\right) \left(\frac{r_m/R_{\rm ns}}{100}\right)^{-1} \, .
\end{equation}

\subsection{Physical assumptions}
\label{sec:GLmodel}
When an accretion disk feeds a magnetised NS, the stellar magnetic
field is expected to thread the disk with field lines, which are 
twisted by the differential rotation in the (Keplerian) disk.  This
interaction between magnetic field lines and disk plasma allows a
continuous exchange of angular momentum between the NS and the disk, via
Maxwell stresses associated to twisted field lines. This couples the evolution of
the NS spin and of the mass accretion rate (GL).

Maxwell stresses are strongest in the inner disk, and become the main
mechanism for angular momentum removal inside the ``magnetospheric
radius", which is defined as $r_m = \zeta \left[\mu^4/ (2 GM \dot{M}^2)\right]^{1/7} $.
The magnetic dipole moment of the NS is $\mu = B_{\rm p} R^3_{\rm ns}/2$, 
in terms of the magnetic field strength at the magnetic pole, B$_{\rm p}$. We will adopt
the normalisation (in c.g.s. units) $\mu_{30} = 0.5~{\rm B}_{\rm p,12}$, where $Q_x \equiv Q/10^x$. At $r_m$ the disk is truncated and the plasma must flow along magnetic
field lines. In the GL model $\zeta \approx 0.52$ and, in general,
$0.3 < \zeta <1 $ (Psaltis \& Chakrabarty 1999).

Accretion can only occur if the inner disk rotates faster than the local
magnetic field lines co-rotating with the NS, i.e. if
$\Omega_K(r_m) > \Omega_{\rm spin}$. Defining the co-rotation radius
$r_{\rm cor}$ as the point where $\Omega_K(r_{\rm cor}) = \Omega_{\rm
  spin}$, the condition for accretion becomes $r_m < r_{\rm cor}$. In
this case the NS accretes the specific angular momentum of matter at $r_m$, thus
receiving a spin-up torque N$_{\rm acc} = \dot{M} (GM r_m)^{1/2}$, i.e. the 
accretion-induced spin-up torque {\it in the absence of threading}.

The outer parts of the disk, at $r> r_{\rm cor}$, rotate slower than the
NS thus extracting its angular momentum as they drag the more
rapidly-rotating magnetic field lines. This region applies a spindown
torque on the NS, with an amplitude that depends on the microphysics
of the interaction between disk and magnetic field.

The {\it total} torque on the NS is thus the sum of these two opposing torques. 
The resulting spin up is in general less than for a non-threaded disk,
at a given accretion rate, due to the negative contribution of regions where 
$r_{\rm cor} < r < r_{\rm out}$. We can formally write
\begin{equation}
\label{eq:def-total-torque}
N_{\rm tot} = N_{\rm acc} \hat{n} \, ,
\end{equation}
where magnetic threading is entirely encoded in the ``torque function" $\hat{n}$ whose value is,
in general, smaller than unity. 

\subsection{The torque function}
\label{sec:torque-function}
As shown by GL, the torque function depends on the relevant radii,
$r_m$ and $r_{\rm cor}$, only through their ratio. Defining the {\it fastness parameter} 
$\omega_s = \Omega_{\rm spin} / \Omega_{\rm K} (r_m) = (r_m / r_{\rm cor})^{3/2}$, 
one can write  $n = \hat{n}(\omega_s)$.

In general, $n(\omega_s)$ should be a decreasing function of
$\omega_s$, reaching zero at some critical value $\omega_c$. If
$\omega_s < \omega_c$ then the torque function is positive, while $n(\omega_s) < 0$
if the fastness parameter exceeds the critical value $\omega_c$. 
The exact shape of $n(\omega_s)$ depends on several details of the disk-magnetosphere 
interaction and can vary among different models (e.g., Wang 1995, 1997; Yi et al. 1997;
Erkut et al. 2005).  On quantitative grounds, however, the main conclusions depend only weakly 
on such details.  We will therefore adopt the GL model from here on to keep the discussion 
focused, and discuss later ways to distinguish the alternatives.

An analytic approximation for the torque function in this model is (GL79b)
\begin{equation}
\label{eq:def-enne}
n(\omega_s) \approx 1.39 ~\frac{1-\omega_s \left[4.03(1-\omega_s)^{0.173} - 0.878\right]}{1- \omega_s} \, ,
\end{equation}
which goes to zero for $\omega_c \approx 0.35$. When this condition is
met, for $r_m \approx 0.5~r_{\rm cor}$, the total torque is zero
and the NS accretes at constant spin. At a fixed magnetic field, a
further decrease in the mass accretion rate would increase $r_m$, in turn increasing
$\omega_s$ and causing $n(\omega_s)$ to become negative: the NS would spin
down while accreting.

\begin{figure}
\centerline{
\includegraphics[width=8cm]{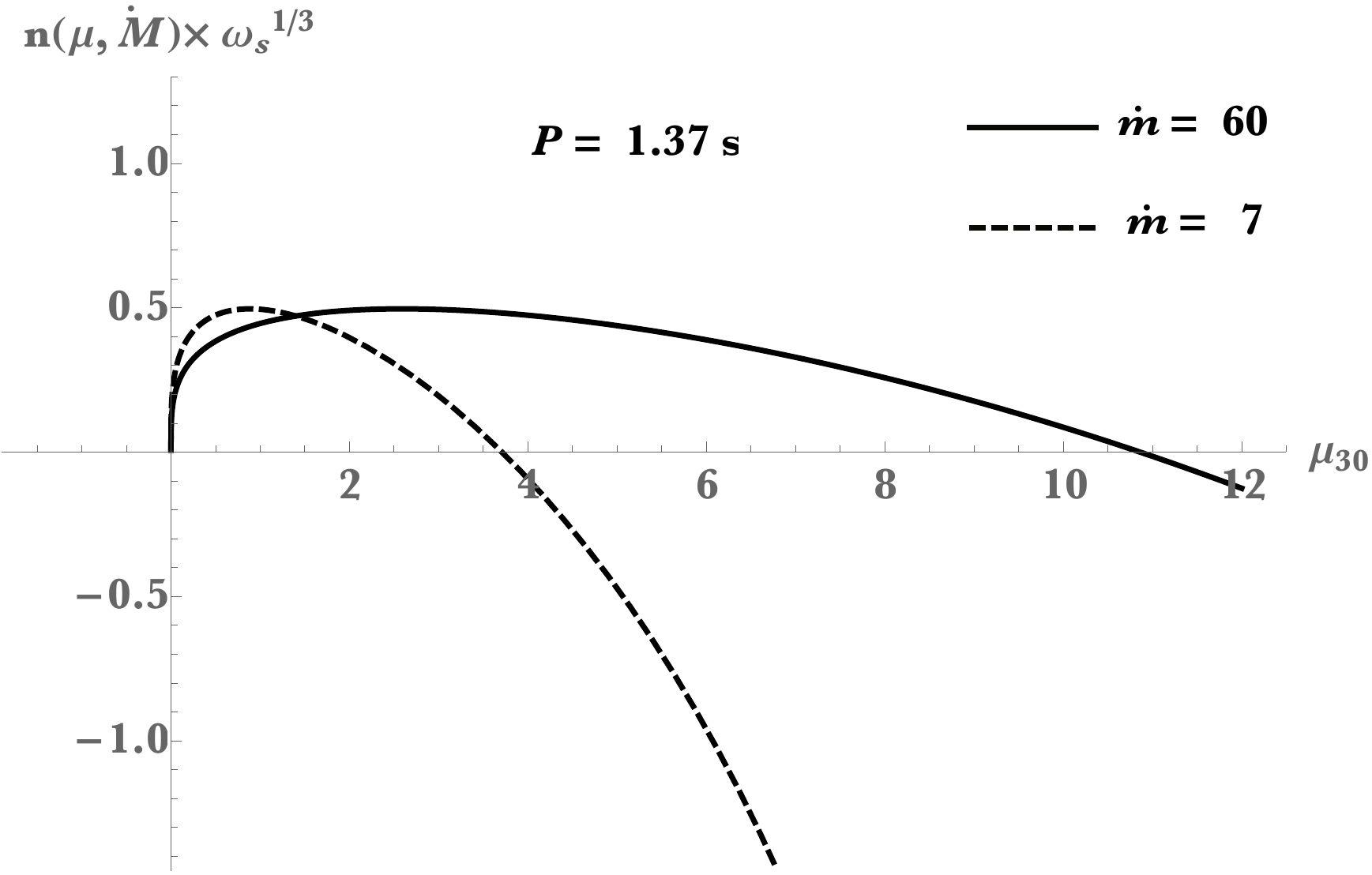}}
\caption{The product (see text) $\omega^{1/3}_s n(\omega_s)$ vs. the
  magnetic dipole moment, $\mu$, for two selected values of $\dot{M}$ in the
  GL model, adopting the spin period P=1.37 s of \nustar. When the
  function is positive, hence N$_{\rm tot} >0$, two possible $\mu$'s exist for 
  each value of $\omega^{1/3}_s n(\omega_s)$.  For
  negative values of the function the degeneracy is broken, and only
  one value of $\mu$ is possible.}
\vspace{-0.1in}
\label{fig:enne}
\end{figure}

We write the total torque as 
\begin{equation}
N_{\rm tot} = N_{\rm 0} \,n(\omega_s)\,\omega_s^{1/3}, 
\label{eq:torque}
\end{equation}
where N$_{\rm 0} = \dot{M} (GM r_{\rm co})^{1/2}$.  The function
$n(\omega_s)\omega_s^{1/3}$ is displayed in Fig.\ref{fig:enne} vs. the
dipole moment
$\mu_{30}$, for two values of the accretion rate. For each positive
value of the function (and hence of $N_{\rm tot}$), two solutions for
$\mu$ are obtained. Their separation increases for increasing
$\dot{M}$.  On the other hand, when the function (and hence N$_{\rm
  tot}$) is negative, only one value of $\mu$ is allowed.

\subsection{Effect of super-Eddington accretion on the disk structure}
\label{sec:deviations}
In the GL model, the properties of standard, gas pressure-dominated, 
geometrically thin 
accretion disks are assumed (Shakura \& Sunyaev 1979; SS hereafter).
Hence a direct application of the GL model to \nustar, with its
super-critical mass accretion rate, needs to be dealt with care.

Formally, the standard SS model does not apply in the case of 
super-Eddington accretion (Abramowicz 2004 and references therein); 
the so-called ``slim disks" (Abramowicz et al 1988; Beloborodov 1998; 
Sadowski et al 2009, 2011) should be used instead,  which generalise the SS 
model to a wider range of conditions\footnote{Abramowicz (2004) notes that even 
the slim disk can be regarded as an higher-order approximation, in $h/r$, to the 
most general case of the ``polish doughnut" (or fat torus).}.  
The crucial difference is that slim disks become radiatively inefficient 
at some critical radius,
inside which a growing share of the locally-released gravitational
energy is advected with the flow rather than being radiated away.
Beloborodov (1998) provides a simple way to estimate this effect, and
compares detailed solutions of relativistic slim disks with radial
profiles of standard SS disks up to large radii\footnote{Radial
  profiles extend down to $r \sim 10 r_g$ in the recent work by Sadowski et
  al. (2011), much smaller than needed for our discussion here (see
  sec. \ref{sec:application}). Also note that, due to different
  notations, $\dot{m}$ defined here is 4.8 smaller than $\dot{m}_{\rm
    bl}$ of Beloborodov (1998) and 3.3 times larger than $\dot{m}_{\rm
    sd}$ of Sadowski et al. (2011).}.  The largest deviations from the
SS disks occur\footnote{assuming a viscosity parameter $\alpha \sim
  0.1$.} at $r \sim (20-25) r_g$, i.e. $\sim 5\times 10^6$ cm for a
NS, in the case $\dot{m}_{\rm bl} \sim$ 100. Here the disk inflates
significantly, and the rotation becomes sub-keplerian by $\sim
(5-10)$\%, in agreement with the results of Sadowski et al. (2011)
for the case $\dot{m}_{\rm sd} \sim 10$. At $r=r_m \sim 10^8$~cm
(which is relevant to our specific situation, see. Sec.4.3), the
deviation from Keplerian rotation is tiny, as are the deviations of
all relevant quantities from the SS model.  We conclude that the
accretion disk is well described by the SS model {if its inner radius
  $r_m \gtrsim $ several $\times 10^7$~cm}. \\

In addition, the disk is likely to remain slightly radiation-pressure-dominated out
to large radii, $r\gtrsim 10^8$ cm. 
Radiation pressure can introduce changes in the disk's density and height profile compared 
to standard GL-type models, and in the exact dependence of $r_m$ on $\dot{M}$ and $\mu$ 
 (Ghosh 1996).  While a quantitative analysis is beyond our scope here, in sec. \ref{sec:refinements} we provide 
 a qualitative discussion of these effects, along with other sources of uncertainty in models for 
 the disk-magnetic field coupling.
\begin{figure}
\centerline{
\includegraphics[width=8.5cm]{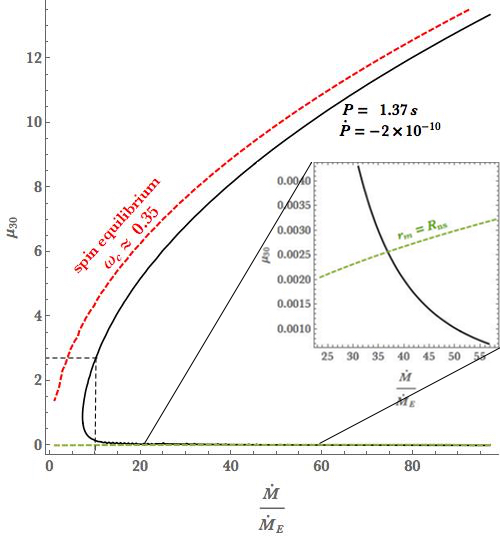}}
\caption{The locus of points in the $\mu-\dot{M}$ plane satisfying
  Eq. (\ref{eq:measured}) for the averaged value of the measured
  $\dot{P}$.  The equilibrium spin condition of the GL model is shown
  as the red, dashed line.  Three main regimes are identified: 1)
  low-$B$ case, $\sim 5 \times 10^{9}-10^{11}$~G, for mass accretion rates
  $10 \lesssim \dot{m} \lesssim 37$; 2) $B$-fields $\sim 10^{13}$~G, for $10 \lesssim \dot{m} \lesssim
  60$; 3) Intermediate values of  $B \sim
  10^{11}-10^{12}$~G only for a narrow range of mass accretion rates around
  the minimum, $7 \lesssim \dot{m} \lesssim 10$. The inset highlights
  the region (above the green dashed line) for which  $r_m > R_{\rm ns}$ 
 and hence pulsations are possible.}
\vspace{-0.2in}
\label{fig:ulx-average}
\end{figure}

\section{Application to 
\titolo: allowed magnetic field values}
\label{sec:application}

Let us now focus our discussion on the ULX \nustar, whose pulsations at $\sim 1.37$ s 
have been recently discovered. In order to characterise the main features 
of the solution for this source, we first consider the $\dot{P} \approx -2 \times
10^{-10}$~s~s$^{-1}$ reported by Bachetti et al. (2014) as representative of the average
over different measurements. In $\S$ \ref{sec:three-regimes} we will exploit the four different values of $\dot{P}$
measured by Bachetti et al. (2014) to further constrain the allowed range of parameters. 

The total torque N$_{\rm tot}$ is related to the instantaneous period derivative,
\begin{equation}
\label{eq:measured}
{\rm N}_{\rm tot} = n(\omega_s) \omega^{1/3} {\rm N}_0 = 2 \pi I \dot{\nu}\,,
\end{equation}
where $I$ is the NS moment of inertia and $\dot{\nu} \simeq
1.07\times10^{-10}$~s$^{-2}$. A numerical solution to this equation
yields $\omega_s$, which identifies a locus of points in the $\mu$
vs. $\dot{M}$ plane. Fig.~\ref{fig:ulx-average}
illustrates the solution for the average $\dot{P}$ quoted above: for
each value of $\dot{M}$ two different values of the $B$ field are
allowed\footnote{Note that this is true in the threaded disk model,
  but not in a model without magnetic threading of the disk.},
reflecting the property of positive torques discussed in
sec. \ref{sec:torque-function} and Fig. \ref{fig:enne}. 

The inset in the left panel of Fig.~\ref{fig:ulx-average} highlights the ``forbidden" area (below the green line) 
in which $r_m\lesssim R_{\rm ns}$ and the disk would reach the NS surface, thus quenching the possibility
to produce sizeable pulsations from the source.

\subsection{Which magnetic field value?}
We define for convenience three main regions in the solution, identified
by the vertical dashed line in Fig. \ref{fig:ulx-average}
at $\dot{m} = 10$.
The narrow range $7 <\dot{m} < 10$ corresponds to a wide range of
magnetic fields: $10^{11}~{\rm G}~ <~{\rm B}_{\rm p} < 5
\times 10^{12}$~G. This is the ``intermediate-$B$" region, or low-$\dot{m}$
region.

For larger values of $\dot{m}$ we have two clearly distinct branches,
a ``low-$B$" region where $B_{\rm p} \lesssim 10^{11}$~G and a
``high-$B$" region where 5$\times 10^{12}$G $<$ $B_{\rm p} \lesssim
2\times 10^{13}$~G.

The low-$B$ branch is far from spin equilibrium: $r_m < 10^7$ cm is
rather close to the NS and 
spin-down torques are negligible.
The high-$B$ branch is close to spin equilibrium, with the disk
truncated very far from the NS ($r_m/R_{\rm ns} \sim 80-90$): a
significant part of the disk is beyond $r_{\rm cor}$ and produces a
spin-down torque.  The intermediate-$B$ solution is not much farther
from spin equilibrium, at least for $B_{\rm p} \gtrsim 10^{12}$~G, yet
this difference will prove to be crucial.

Further inspection of Fig.~\ref{fig:ulx-average} shows that:\\ {\em
  (a)} the low-$B$ region exists only if\footnote{Note that the exact
  numerical value of this lower limit for $B_{\rm p}$ is
  specific to the GL model.} $B_{\rm p}\gtrsim 5\times 10^9$~G, or
else the disk reaches the NS surface, quenching the pulsations. \\ 
{(b)} The $\mu-\dot{M}$ curve turns around at $\dot{M}
\sim 7 \,\dot{M_E}$, corresponding to $\mu_{30} \sim 1$, or $B_{\rm p}
\sim 2\times 10^{12}$ G. Eq.~(\ref{eq:mdot}) thus implies that $b
\gtrsim 0.12$, indicating that the source must be mildly 
beamed.  Even in the most favorable case, its
actual X-ray luminosity is L$_X \gtrsim 10^{39}$~erg~s$^{-1}$, which has
important implications (\S4.3.1).
\begin{figure*}
\centerline{
\includegraphics[width=8cm]{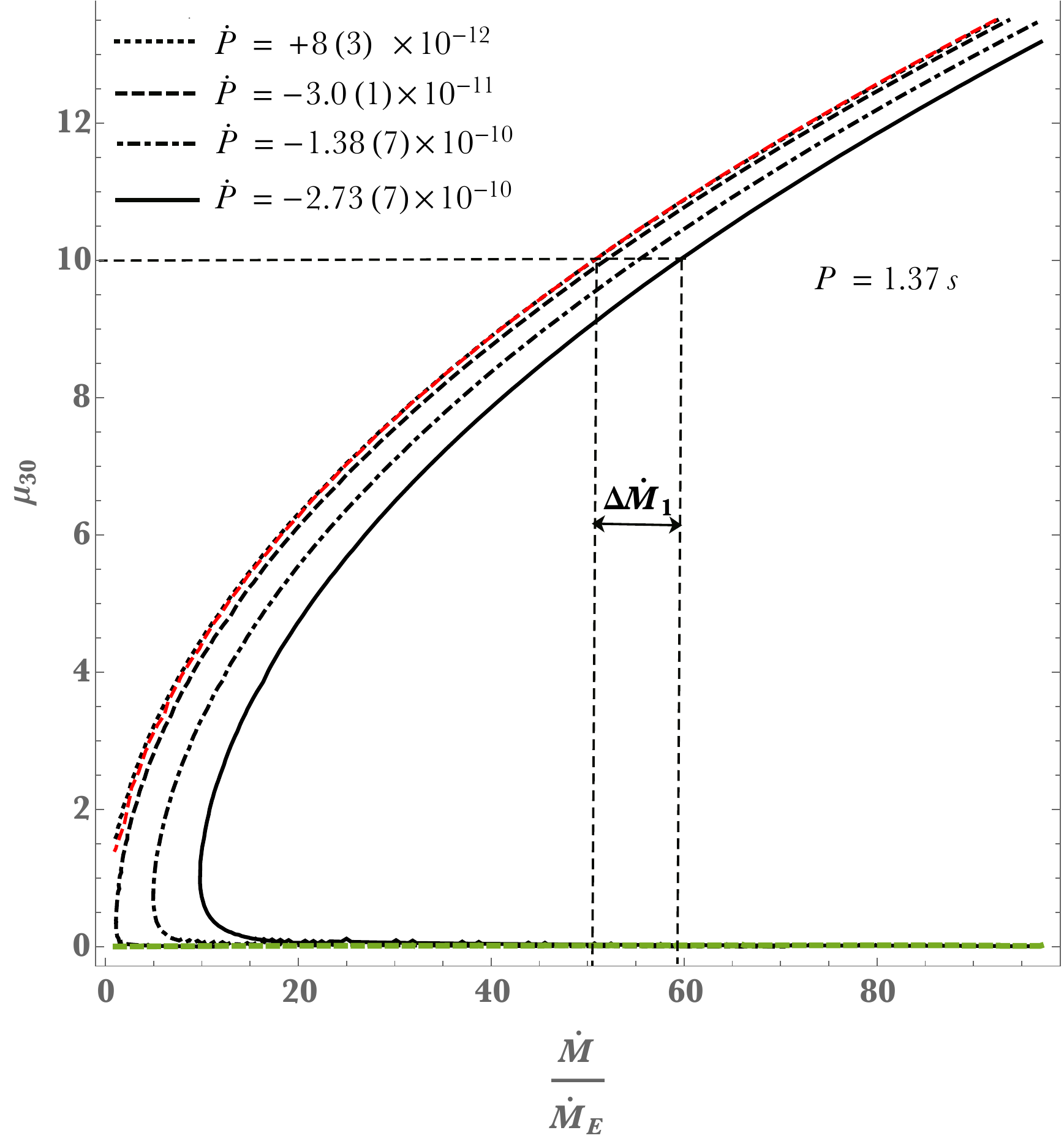}
\hspace{0.4in}
\includegraphics[width=8cm]{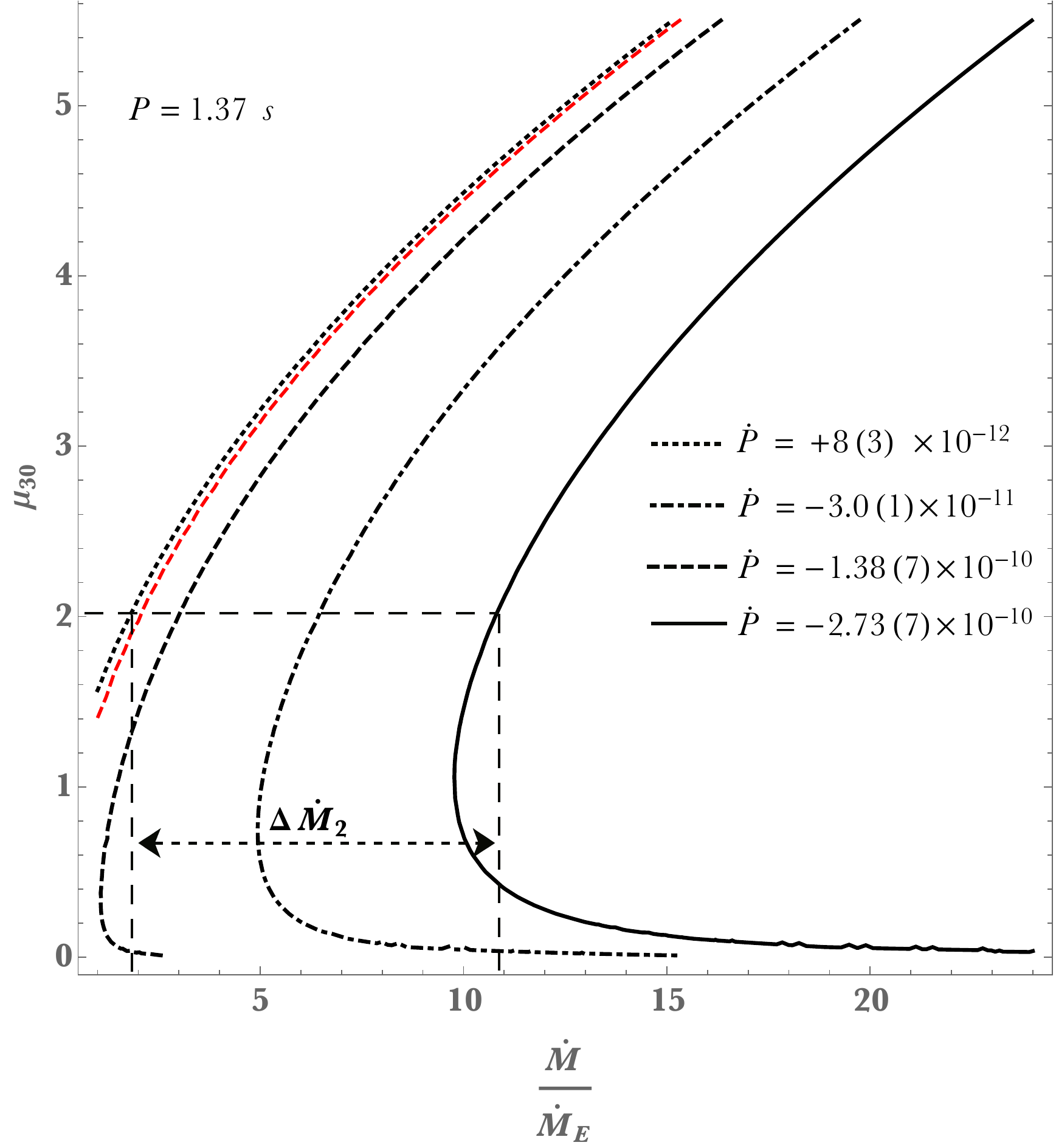}}
\caption{Same as in Fig.~\ref{fig:ulx-average}, but using the four
  different values of $\dot{P}$ reported by Bachetti et al. (2014) in
  four distinct observations. The red dashed line indicates 
  the equilibrium condition N$_{\rm tot} = \dot{\nu} = 0$.  Note that 
  no solution exists in the low-$B$ branch for the equilibrium case or the positive 
  $\dot{P}$. The vertical dashed lines indicate a representative case for
  the variations in mass accretion rate needed to reproduce the
  observed $\dot{P}$ fluctuations. {\em Left panel:} the complete
  solution highlighting the high-$B$ branch, with $\Delta
  \dot{M}/\dot{M} \sim 20$ \%; {\em Right panel:} the intermediate-$B$
  region with $\Delta \dot{M}/\dot{M} \sim 5$.}
\vspace{-0.2in}
\label{fig:four-pdots}
\end{figure*}
However, before analyzing each solution in detail, we must discuss an
additional piece of information that comes from the observations.
\subsection{$\dot{P}$ fluctuations}
\label{sec:fluctuations}
Table 3 of Bachetti et al. (2014) lists variations in the
measured $\dot{P}$  of up to one order of magnitude at different epochs
(see caption of Fig. \ref{fig:four-pdots} for the individual values).
A change in the sign of $\dot{P}$ (spin-down) at one particular epoch
(obs. 007) may have also been seen.
%

We note that the variations of the pulse
phase/shape correlated with variations in the mass accretion rate
may mimick the effects of spin period derivatives, potentially
affecting the results of phase-coherent timing analyses. 
The possibility of such mimicking effects was discussed in 
relation to accreting millisecond pulsars in transient low mass X-ray 
binaries, where the pulsations could be studied over a source flux decrease
of typically one order of magnitude on a timescale of tens of
days  (Patruno et al. 2009; Patruno 2010). 
In order to interpret the measured period derivatives 
$\dot{\nu} \sim 10^{-13}$ s$^{-2}$ in these systems, 
accretion-rate induced changes in the pulse profile would be required 
which can cause a cumulative phase shifts of $\sim 0.2-0.5$ cycles. 

In the case of \nustar, the observations in which pulsations were detected had durations 
of a few days and displayed only modest flux variations of up to a few percent
in individual observations and of up to $\lesssim$ 60\% across different observations
(we note this is a broad upper limit, since some of these variations have been 
attributed to contamination by the nearby ULX source M82 X-1; Bachetti et al. 2014). 
Even if phase-shifts as large as $\sim 0.2-0.5$ cycles were produced in \nustar~ by 
pulse profile changes over a 5-day observation, a $\dot{\nu} \sim {\rm~a few} \times 10^{-12}$ 
s$^{-2}$ would result, comparable to the largest measurement 
errors, e.g. those of obs. 007. 
We conclude that the measured $\dot{P}$ variations 
observed in \nustar~ are intrinsic to the source and amount to at least 
an order of magnitude for only modest variations of the source flux. 
For the sake of definiteness, we will therefore adopt in the following the values 
reported by Bachetti et al. (2014), keeping in mind that the positive $\dot{P}$ (spin-down) 
measurement (obs. 007) might not be significant, in consideration of its error bar 
and the discussion above. 

In the following, we apply condition (\ref{eq:measured}) to
each of the four $\dot{P}$ values reported by Bachetti et al. (2014),
and show the corresponding 
solutions in
Fig. \ref{fig:four-pdots}.  These will provide 
additional constraints on the characteristics of the source. 

\subsection{The three possible regimes}
\label{sec:three-regimes}
We discuss  here the relative merits of the different solutions and constrain
the physical properties of the hyper-accreting NS in \nustar.

\subsubsection{Low-B ($5 \times 10^9$G $<$ B$_{\rm p}  \lesssim 10^{11}$G)}
\label{sec:low}
This solution corresponds to $10 < \dot{m} \lesssim 37$, hence a
highly super-Eddington luminosity L$_X > 2\times 10^{39}$ erg
s$^{-1}$, compatible with a modest amount of beaming $b>0.2$. The
inner disk radius ranges from $r_m \gtrsim$ R$_{\rm ns}$ to $r_m\simeq
8$ R$_{\rm ns}$.

With the inner edge of the disk close to the NS, magnetic threading is
essentially irrelevant. The NS is only subject to the material torque,
N$_{\rm acc} \approx \dot{M} (G M r_m)^{1/2} \propto \dot{M}^{6/7}$.
Fluctuations of $\dot{P}$ by one order of magnitude can only be
explained by changes in $\dot{M}$ of comparable amplitude, in stark contrast
with the relatively stable X-ray emission of the source during the
same period. Furthermore, the brief phase of torque reversal 
(or equivalently very low torque) revealed
during obs. 007 would be impossible in these conditions
(see the left panel of Fig.~\ref{fig:four-pdots}). 


Moroever, since $B \sim 10^{11} \ll 10^{13}$~G,  it would very hard to exceed 
L$_X = 2 \times 10^{39}$~erg~s$^{-1}$, even for an optically thick accretion
column (Basko \&
Sunyaev 1976a)

\subsubsection{Intermediate-B ($10^{11}$G $<$B$_{\rm p} < 5 \times 10^{12}$G)}
\label{sec:inter}
This case corresponds to mass accretion rates in a narrow range, $\dot{m} \sim 7-10$ (Fig~\ref{fig:ulx-average}), 
which in turn implies a significant beaming of the emitted radiation (eq. \ref{eq:mdot}), 
$0.12 \lesssim b \leq 0.17$, and a minimal source
luminosity, L$_X \sim (1-2) \times 10^{39}$~erg~s$^{-1}$, 
marginally consistent with the theoretical maximum for an optically thick accretion column (Basko \& Sunyaev 1976a).

Similarly to the low-$B$ case, this regime cannot explain the large $\dot{P}$ oscillations.
Solutions for all four measured values of $\dot{P}$ only exist for $\mu_{30} > 1$, or B$_{\rm p} > 2\times 10^{12}$ G 
(Fig.~\ref{fig:four-pdots}, right panel). Even in that case, however, fluctuations of 
$\dot{M}$ by a factor of $\sim 3-5$ would be required to reproduce the observations (as indicated by $\Delta \dot{M}_2$ 
in Fig.~\ref{fig:four-pdots}, right panel), in contradiction with the fairly stable source emission.

We note also that a beaming factor in 
$b \sim 0.1-0.15$ is not easily achieved even in the
optically thick environment of sources accreting at $\gtrsim
\dot{M}_E$ (Basko \& Sunyaev 1976b): \nustar~would thus represent an
extreme example of this issue. 

\subsubsection{High-B ($5 \times 10^{12}$G $< $B$_{\rm p} \lesssim 2\times10^{13}$G)}
\label{sec:high}
This solution corresponds to 10 $< \dot{m} < 60$, the upper bound
being the accretion rate that gives the luminosity $L_{\rm X, iso}
\sim 10^{40}$~erg~s$^{-1}$ for a negligible beaming ($b \sim 1$,
Eq. \ref{eq:mdot}).  On the upper branch in
Fig.~\ref{fig:ulx-average} the implied
$B$-field is in the range $(0.5-2) \times 10^{13}$ G. The inner disk
radius is $r_m \sim (80-90)$ R$_{\rm ns}$, the fastness parameter
$\omega_s \approx (0.25-0.31)$ and the disk luminosity $L_{\rm disk}
\sim (0.06-0.3) {\rm L}_{\rm E}$.  At this distance from the NS the
disk has a small thickness $h(r_m)/r_m \lesssim 0.1$ and negligible
deviations from the SS model (Beloborodov 1998).

Since the NS is close to spin equilibrium, relatively small variations
of $\dot{M}$ can explain the large variations in $\dot{P}$. Required $\dot{M}$ 
variation get progresively smaller for higher magnetic fields: the left
panel of Fig.~\ref{fig:four-pdots} shows that, for $\mu_{30}$ = 10,
values of $\dot{m} \sim (50-60)$ can reproduce all the observed torque
variations, including the torque reversal episode.

From an independent argument, the isotropic X-ray luminosity
$L_{\rm X, iso} \sim 10^{40}$~erg~s$^{-1}$ can be attributed to a
reduction in the electron scattering cross-section, 
caused by the strong $B$-field (Basko \& Sunyaev 1976a, Paczynski 1992). 
The scattering cross-section scales as\footnote{The dependence on the direction of propagation is
neglected here for simplicity.} $(E_{\gamma}/E_{\rm cyc})^2$, where 
$E_{\rm \gamma}$ is the photon energy and $E_{\rm cyc} \simeq 11.6 ~B_{\rm p,12}$ keV is 
the cyclotron energy. For photons at $E_{\gamma} \sim 30$ keV, 
a $B$-field $\sim 10^{13}$ G would be required to reduce the cross-section by a factor $\sim 50$. 
%
For the maximum luminosity from a plasma which is optically thick 
to electron scattering, and integrating over photon energy and propagation angle, 
Paczynski (1992) derived,  
\begin{equation}
\label{eq:pac92}
L_{\rm E, mag} = 2 L_{\rm cr} \left(\frac{B_{\rm p}}{10^{12}{\rm G}}\right)^{4/3} \, ,
\end{equation}
where L$_{\rm cr}$ represents the limiting luminosity in the absence
of magnetic fields. Substituting L$_{\rm E}$ and requiring that
L$_{\rm E, mag} = L_{\rm X, iso} = 10^{40}$ erg s$^{-1}$, gives
$B_{\rm p} \sim 1.5 \times 10^{13}$ G,  in line with the value obtained from the torque analysis. 
As discussed above, the geometry of an accretion column can increase L$_{\rm cr}$ a few
times above L$_{\rm E}$, thus reducing the requirement on $B_{\rm p}$. For a maximum 
L$_{\rm cr} \sim 10^{39}$~erg~s$^{-1}$ we would still need $B_{\rm p} \sim 5\times 10^{12}$ 
according to Eq.(\ref{eq:pac92}).

The above arguments point independently to the same range of
values of the $B$-field, $5\times 10^{12}$~G $\lesssim B \lesssim 2\times
10^{13}$~G, and to only a modest degree of beaming ($0.5\lesssim b<1$).
Therefore, the high-$B$ solution provides a satisfactory
intepretation of the source properties.

\subsection{Alternative models and refinements}
\label{sec:refinements}
Before concluding our discussion, we should note that alternative
models for the disk-magnetic field interaction lead to slightly
different locations of the inner disk, $r_m$, and a slightly different
dependence of the torque function $n(\omega_s)$ on the fastness
parameter. Adopting e.g., the model by Yi et al. (1995;  also discussed
by Andersson et al. 2005), where $\omega_c \approx 0.71$, the low-$B$
branch is unaffected by the change while the high-$B$ branch - which is the
most affected - is shifted upwards by $\lesssim30$ \% at most. 

This is to be expected
given that the low-$B$ solution is far from the equilibrium condition (N$_{\rm tot}=0$), 
and thus it is essentially insensitive to magnetic threading, while the high-$B$ solution 
is most sensitive to the details of torque balance, being close to spin equilibrium.
This also offers a key to understand the impact of radiation-pressure
in GL-type models. The low-$B$ branch would still be mostly unaffected
because, as stated before, it is insensitive to the details of magnetic threading.
For $B_{\rm p} > 10^{12}$ G (including both the intermediate-$B$ and the high-$B$ case) 
the inner disk will be slightly radiation pressure dominated ($p_{\rm rad}/p_{\rm gas} \sim 
1.5-4$ for $\dot{m} = (10-50)$, see eq. 5.56 in Frank, King, Raine 2002). 
An in-depth study of this issue will be warranted
in the future.  

\subsection{Comparing with results from recent studies}
To further clarify our discussion we briefly compare our results with those recently presented in the literature. 
In the discovery paper by Bachetti et al. (2014) the condition of spin equilibrium was assumed and translated
into the simple equality $r_{\rm co} = r_m$. Assuming a purely material torque, this was used in combination with the
average value of $\dot{P}$ to constrain the accretion rate. However, as noted by the same authors, the resulting value of 
$\dot{M}$ falls short of what required to explain the source luminosity.  In order to alleviate this tension, Tong (2014) 
discussed possible caveats such as the combination of a moderate beaming, a large NS mass and a complex magnetic 
field structure. 

Eksi et al. (2014) and Lyutikov (2014), on the other hand, considered a magnetically-threaded disk and, still assuming the 
source to be close to spin equilibrium, estimated the NS field to be $\sim 10^{14}$ G and $10^{13}$~G, respectively. 
The difference in these values  comes largely from the different source luminosity that these authors adopted. This range of 
$B$-fields corresponds to our ``high-$B$" solution (sec. \ref{sec:high}).

Lasota \& Kluzniak (2014) noted that the equilibrium condition need not be verified, and discuss a solution 
in which the NS has a low magnetic field ($\sim 10^9$ G) and accretion rate $\dot{M} \sim 50 \dot{M}_{\rm E}$. This corresponds
to our ``low-B" solution (sec. \ref{sec:low}).

In our work we considered a magnetically-threaded disk without prior assumptions on how close the system is to spin equilibrium.
By using the average value of $\dot{P}$ we have shown that a continuum of solutions in the $B$ vs. $\dot{M}$ plane is allowed, 
and the cases discussed above represent specific examples, as already noted. 

The key addition in our study is that we took into account the large fluctuations in the measured values of $\dot{P}$, which are associated 
with small fluctuations of the source luminosity, as found by Bachetti et al. (2014). Unless variations of the beaming factor are invoked to 
keep the luminosity stable in spite of large variations in $\dot{M}$, this behaviour is compatible only with the NS
being close to the spin equilibrium condition. 
This singles out the high-$B$ solution, with a preferred value of $\sim 10^{13}$ G which is lower than the one derived by Eksi et al. (2014), and agrees
with the estimate of Lyutikov (2014). Such a field is strong enough to affect radiative transfer in the emission region and allow a significantly super-Eddington 
luminosity. Further implications of this scenario are briefly commented in the next section.

\section{Summary and discussion}
Our work has provided strong arguments in favor of the interpretation of
\nustar~as a highly magnetized NS with a magnetic field $\sim
10^{13}$~G, a mass accretion rate $\sim (20-50) \dot{M}_{\rm E}$
corresponding to an accretion luminosity $\sim (0.4-1) \times 10^{40}$
erg s$^{-1}$, and little beaming, if at all ($0.5 \lesssim b < 1$).  
In the following we briefly discuss further details of this scenario and some
of its main implications.

The high magnetic field keeps the disk edge far from the NS surface,
at $\lesssim (80-90)$ R$_{\rm ns}$, 
%
and helps maintaining the large
luminosity at the base of the accretion column.  
For a simple dipole
geometry, the field has decreased by a factor $\sim 5$ at a
few kilometers above the NS surface. Hence, the infalling matter in the
accretion column is subject to a super-critical flux of radiation
already in the close vicinity of the NS. The large optical thickness
of the flow, however, shields most of the material from this
radiation, thus preserving the stability of the flow. To be specific,
we will assume an emitting region close to the base of the accretion column
of cross-sectional area $A_{\rm em} \sim 10^{11}$ cm$^2$, as required
for thermal radiation at the observed luminosity to peak around/close
to the spectral range in which \nustar~has been observed ($<$ 50 keV).
Let us consider $\dot{m} = 20$ and freely falling matter with $v_{\rm
  ff} \sim 0.3 c$ at a few kilometres from the NS surface. The
particle density in the flow is $n\sim \dot{M}/(v_{\rm ff} A_{\rm em}m_p)
\sim 10^{22}(\dot{m}/20)A^{-1}_{\rm em,11}$~cm$^{-3}$ which implies $\tau = n \sigma_T l \sim 500
(\dot{m}/20) (l_5/A_{\rm em,11}) $ where, to be conservative, we
allowed for an elongated shape of the flow ($l < \sqrt{A_{\rm em}}$,
cf. Basko \& Sunyaev 1976a), and assumed a relatively low value of
$\dot{m}$.  Radiation from below will continuously ablate the outer
layers of the accretion column to a scattering depth $\tau \sim$ a few 
and, even if the particles were flung away close to the speed of light,
the resulting outflow could only amount to a tiny fraction of the
inflow $\sim \tau^{-1} c/ v_{\rm ff} \sim 0.007$: the material is
easily replenished and accretion can proceed steadily.

Super-Eddington irradiation of the disk from the hot spot on the NS 
surface might affect the disk structure close to the inner boundary. 
By a similar argument to the one above, we conclude that 
shielding of radiation remains effective even at the inner edge of the disk 
where $r_m \sim 10^8$ cm and $h(r_m)/r_m \sim 0.1$ , for a wide range of conditions. 
The optical thickness of the flow at the inner disk boundary, where the disk 
is also geometrically thicker,  guarantees that the disk will survive 
the large radiation produced at the base of the accretion column. 

Observations of the source prior to the latest ones were reported
by Feng \& Karet (2007) and Kong et al.  (2007). The source was found
to alternate between very bright states, with L$_X \sim (0.7-1.3)
\times 10^{40}$ erg s$^{-1}$, and dim undetected ``off" states, below
the sensitivity level of the Chandra observations (this corresponds to 
a luminosity of $\sim 2.5\times 10^{38}$~erg~s$^{-1}$).  Given
that the system is close to the condition for spin
equilibrium\footnote{For a constant mass accretion rate} ($r_m \simeq
0.5~r_{\rm cor} $ in the GL model, and $\sim 0.8-0.9$ in alternative
models), a relatively small decrease in the accretion rate can cause a
transition to the propeller regime, in turn causing a 
sudden decrease of the source luminosity.  In fact in this case 
accretion onto the neutron star would be mostly inhibited, and the 
source luminosity may be dominated by the emission from the disk , 
which is truncated at $r_m \gtrsim r_{\rm cor} \simeq 2\times 10^8$
cm. Eq.~(\ref{eq:disc-luminosity}). We note that the expected 
disk luminosity
$L_{\rm disk}\sim 0.15 L_{\rm E}\sim 2\times 10^{37}$~erg~s$^{-1}$,
would well below present upper limits to luminosity during 
the source "off" state. Since the magnetospheric radius
scales as $\dot{m}^{-2/7}$, a decrease of $\dot{m}$ by a factor $\sim
10$ with respect to the level at which the source emits at 
L$_X \sim 10^{40}$ erg s$^{-1}$ would be needed in order 
to cause the transition to the propeller regime in the GL model. 
A decrease
by only a factor $\gtrsim 2$ would suffice for, e.g. an equilibrium
condition with $r_m = 0.8~r_{\rm cor}$, as envisaged in other 
disk-magnetosphere interaction models. The expected transition luminosity is
$\sim 10^{39}$~erg~s$^{-1}$ and $5\times 10^{39}$~erg~s$^{-1}$,
respectively, in the two cases.

Note that, if the magnetic field is significantly lower, the inner edge of the disk
would extend closer to the NS and larger decreases in
$\dot{m}$ would be required for the propeller mechanism to set in. 
Correspondingly the source would visible over a wider range,
before the suddenly switchinh to much lower luminosities.
Hence we argue that all flux levels so far measured in 
\nustar are consistent with the scenario involving of a $\sim 10^{13}$~G NS. 

The X-ray spectral characteristics of the source can provide
additional clues on the B-field strength of \nustar. The power law-like X-ray
spectrum of many accreting X-ray pulsars was found to bend steeply
above a cutoff energy, E$_{\rm cut}$, which is empirically related to
the energy of cyclotron resonance features through E$_{\rm cyc} ~ (1.2
- 2.5) $E$_{\rm cut}$ (Makishima et al 1990).  If the same
relationship holds for \nustar, its X-ray spectrum would be expected not
to show any cutoff up to energies of at least $\sim$~50 keV, if
B$\gtrsim 10^{13}$ G. On the other hand, a solution with B$\lesssim 5
\times10^{12}$ G would be characterised by a spectral cutoff at
energies\footnote{We caution however that 
for very high accretion rates cyclotron-related spectral 
features might shift to lower energies, as observed e.g. in the 
highest luminosity states of 4U0115+63 and V0332+53 (Nakajima et al. 2006; 
Tsygankov et al. 2006). This might be due 
to an increased height of the post-shock 
region of the accretion column.} $\lesssim (20-40)$ keV. 

We finally note that the almost circular orbit of the system 
($e \lesssim 0.003$; Bachetti et al. 2014), implies a long lasting action of tidal interactions
after the supernova explosion that accompanied the NS birth. This was most likely achieved 
during the Roche-lobe overflow phase, or in the phase immediately preceeding it, characterized 
by very short circularization timescales (Verbunt \& Phinney 1995, Belczynski et al. 2008). Under 
these circumstances it is not possible to use the low eccentricity alone to constrain the NS age. 
However, if it were possible to constrain the age of the system through, e.g. population synthesis 
studies, then this would offer the intriguing possibility to also test theories of magnetic field 
evolution in highly magnetised NSs. 

In fact, detailed magnetothermal simulations (Vigan\'o et al. 2013) show that the
lifetime of the NS dipole field strongly depends on where the field
is located, and on the presence or not of a strong toroidal component in addition to the dipole. 
Crustal fields tend to decay faster than those
rooted in the NS core, due to the different electrical conductivities, and additional toroidal 
components increase the rate of dissipation, making the dipolar field decay faster than in 
cases with no toroidal field. Internal fields that are mostly concentrated in the (superconducting) 
core have the slowest rate of dissipation.
Therefore, the lifetime of a $\sim 10^{13}$ G field could be $\lesssim $ 1~Myr if it were 
rooted in the NS crust, and coupled to a strong toroidal component, or $\gtrsim 10^8$ yrs 
if the field were largely confined to the core.

\section*{Acknowledgements}
For this work S.D. was supported by the SFB/Transregio 7, funded by the 
Deutsche Forschungsgemeinschaft (DFG).
RP was partially supported by NSF grant No. AST~1414246 and
Chandra grants (awarded by SAO) G03-13068A and G04-15068X (RP).
LS acknowledges discussions with P.G. Casella, G.L. Israel and A. Papitto.
LS acknowledges partial support by PRIN INAF 2011.\\

\end{document}